\documentclass[twocolumn]{autart}
\usepackage{mathrsfs}    

\usepackage{graphicx}          
\usepackage{amsmath,amssymb,color}
\usepackage{dsfont}

\begin{document}

\begin{frontmatter}

\title{On Matrix Method of Symmetric Games\thanksref{footnoteinfo}} 

\thanks[footnoteinfo]{This paper was not presented at any IFAC
meeting. Corresponding author J.~Zhu. Tel. +86-13851781823.}

\author[NNU]{Lei Wang}\ead{wanglei1010621@163.com},
\author[WFU]{Xinyun Liu}\ead{liuxinyun1224@163.com},
\author[NNU]{Ting Li}\ead{dylliting@163.com},
\author[NNU]{Jiandong Zhu*}\ead{zhujiandong@njnu.edu.cn},

%
\address[NNU]{Institute of Mathematics, School of Mathematical Sciences, Nanjing Normal University, Nanjing, 210023, China}             
\address[WFU]{School of Mathematics and Information Sciences, Weifang University, Weifang, 261061, China}


\begin{abstract}
This paper provides a new version of matrix semi-tensor product method based on adjacent transpositions to test symmetric games. The advantage of using adjacent transpositions lies in the great simplification of analysis of symmetric games. By using the new method, new necessary and sufficient conditions for symmetric games are proposed, and a group of bases of a symmetric game space can be easily calculated. Moreover, the testing equations with the minimum number can be concretely determined. Finally, two examples are displayed to show the effectiveness of the proposed method.
\end{abstract}

\end{frontmatter}

\newtheorem{theorem}{Theorem}
\newtheorem{proposition}{Proposition}
\newtheorem{remark}{Remark}
\newtheorem{lemma}{Lemma}
\newtheorem{example}{Example}
\newtheorem{assumption}{Assumption}
\newtheorem{corollary}{Corollary}
\newtheorem{definition}{Definitoin}

\section{Introduction}

Game theory studies mathematical models of strategic interactions among rational agents. Modern game theory was formally established after John von Neumann proved the basic principles of game theory \cite{kuhn1958john,neumann1928theorie}, which has been paid wide attention and applied to many fields such as economics, biology, finance, computer science and so on. The concept of symmetric game was first proposed by von Neumann in \cite{neumann1928theorie}, and the details can be seen in \cite{von1947theory}, which is a game where the payoffs for playing a particular strategy depend only on the other strategies employed, not on who is playing them.
In other words, a symmetric game means that all players have the same set of strategies and fair payoffs.
In real world, symmetric games are universal. For example, the well-known games rock-paper-scissors, prisoner's dilemma are symmetric games. Moreover, the importance of symmetric game lies in the discussion on existence of Nash equilibria \cite{hefti2017equilibria,nash1951non}. Therefore, how to check whether a game is a symmetric game is an important problem. In \cite{cao2018symmetric}, symmetric games are classified into three types, including ordinary symmetric games, renaming symmetric games and name-irrelevant games. The symmetric games discussed in this paper are actually ordinary symmetric games.

In the past decade, Cheng and his collaborators applied the semi-tensor product (STP) of matrices to game theory \cite{cheng2010analysis,cheng2014evolutionarily}, which enable a game to be expressed as a linear representation. Based on the linear representations of games, some problems about potential games, networked games and evolutionary games have been solved \cite{cheng2014on,cheng2014evolutionarily,liu2016on}. It should be noted that the matrix method based on STP is also powerful for symmetric games. In
\cite{cheng2017linear}, algebraic conditions for symmetric games have been proposed by using the linear representation of symmetric group in structure vectors. Compared with the results in \cite{cheng2017linear},
some simpler algebraic conditions for symmetric games
 are given   in \cite{cheng2018boolean} via some STP operations. Furthermore, for Boolean games,
a group of bases of the symmetric game subspace is constructed via a recursive procedure in \cite{cheng2018boolean}.
For general finite games, construction algorithms of bases of symmetric game subspaces are given in \cite{hao2018on}
and \cite{li2019symmetry}, respectively. It should be noted that the concept of
strategy multiplicity vector and an equivalent condition for symmetric games in \cite{cao2018symmetric} play an important role in \cite{cheng2018boolean} \cite{hao2018on} and \cite{li2019symmetry}.

In this paper, we still use the matrix method based on STP to investigate symmetric games. However, due to the introduction of  adjacent transpositions, some new equivalent conditions for symmetric games are derived, and the proof is greatly simplified via a purely algebraic approach without using the
concept of strategy multiplicity vector \cite{cao2018symmetric}.
Based on our results, a group of bases of the symmetric game subspace can be easily constructed, and the testing equations with the minimum number can be concretely determined.

The rest of this paper is organized as follows: Section 2 gives some preliminaries and problem statement. In Section 3, the main results are presented. Section 4 is a brief conclusion.

\section{Preliminaries and problem statement}

In this section, we give some necessary preliminaries and problem statement. Firstly, we list the following notations.
\begin{itemize}
\item[$\bullet$] $\mathcal{D}=\{0,1\}:$ the set of values of logical variables;
\item[$\bullet$] $\delta_k^i: $ the $i$-th column of $I_k$;
\item[$\bullet$] $\Delta_k:=\{\delta_k^i: i=1,2,\cdots,k\}$;
\item[$\bullet$] $\delta_k[i_1~i_2~\cdots i_n]:=[\delta_k^{i_1}~\delta_k^{i_2}~\cdots~\delta_k^{i_n}]$;
\item[$\bullet$] ${\mathcal{M}}_{m\times n}:$ the set of $m\times n$ matrices;
\item[$\bullet$] ${\mathcal{L}}_{m\times n}:=\{L\in{\mathcal{M}}_{m\times n}|\textrm{Col}(L)\subseteq{\Delta}_m\}$;
\item[$\bullet$] $\ltimes:$ the left semi-tensor product of matrices;
\item[$\bullet$] $\textbf{1}_n$: the $n$-dimensional column vector of ones;
\item[$\bullet$] $0_{m\times n}$: the $m\times n$ matrix with zero entries;
\item[$\bullet$] $\textbf{S}_n$: the $n$-th order symmetric group, i.e., a permutation group that is composed of all the permutations of $n$ things;
\item[$\bullet$] $\mathbb{R}$: the set composed of all the real numbers.

%
%
\end{itemize}

\begin{definition}\cite{MONDERER1996124}
\label{def1}
A finite game is a triple $G=(N, S, C)$, where
\begin{description}
  \item[(1)] $N=\{1, 2, \cdots,n\}$ is the set of $n$ players;
  \item[(2)] $S=S_1\times S_2\times\cdots\times S_n$ is the set of strategy profiles, where $S_i=\{s_1^i, s_2^i, \cdots, s_{k_i}^i\}$ is the set of strategies of player $i$;
  \item[(3)] $C=\{c_1, c_2, \cdots, c_n\}$ is the set of payoff functions, where $c_i: S\rightarrow \mathbb{R}$ is the payoff function of player $i$.
\end{description}
Denote the set composed of all the game above by $\mathcal{G}_{[n;k_1,\cdots, k_n]}$.  When $|S_i|=k$ for each $i=1, 2, \cdots,n$, we denote it by $\mathcal{G}_{[n;k]}$.
\end{definition}

\begin{definition}\cite{cheng2010analysis}
\label{def2}
Let $A\in\mathcal{M}_{m\times n}$, $B\in\mathcal{M}_{p\times q}$. The left semi-tensor product of $A$ and $B$ is defined as
\begin{equation}
\label{eqn1}
A\ltimes B=(A\otimes I_{\frac{\alpha}{n}})(B\otimes I_{\frac{\alpha}{p}}),
\end{equation}
where $\otimes$ is the Kronecker product and $\alpha=\operatorname{lcm}(n,p)$ is the least common multiple of $n$ and $p$. When no confusion may arise it is usually called the semi-tensor product (STP).
\end{definition}
If $n$ and $p$ in Definition $\ref{def2}$ satisfy $n=p$, the STP is reduced to the traditional matrix product. So, the STP is a generalized operation of the  traditional matrix product. Therefore, one can directly write $A\ltimes B$ as $AB$.
%
%


Given a game $G\in\mathcal{G}_{[n;k]}$, by using the semi-tensor product method \cite{cheng2014on}, one can  write each strategy $x_i$ into a vector form $x_i\in\Delta_k$, and express every payoff function $c_i$ as
\begin{equation}
\label{eqn3}
c_i(x_1, x_2, \cdots, x_n)=V_i^c\ltimes_{j=1}^nx_j, ~i=1, 2, \cdots, n,
\end{equation}
where $\ltimes_{j=1}^nx_j\in \Delta_{k^n}$ is called the STP form of the strategy profile, and
$V_i^c$ is called the structure vector of $c_i$.

\begin{definition}
(Definition 2.8 of \cite{cheng2010analysis}) A swap matrix $W_{[m,n]}=(w^{IJ}_{ij})$ is an $mn\times mn$ matrix, defined as follows:
\par
Its rows and columns are labeled by double indices. The columns are arranged
by the ordered multi-index Id$(i_1,i_2;m,n)$, and the rows are arranged by the ordered
multi-index Id$(i_2,i_1; n,m)$. The element at the position with row index $(I,J)$ and column index $(i,j)$ is
$$
w^{IJ}_{ij}=\left\{\begin{array}{l}
                      1,\ \  I=i \mbox{ and } J=j,\\
                      0,\ \  otherwise.
                      \end{array}\right.
$$
\end{definition}
For example, letting $m = 2$, $n = 3$, we get the swap matrix
$$
\begin{array}{rc}
(11)(12)(13)(21) (22) (23)~& \\
W_{[2,3]} = \left[\begin{array}{cccccc}
                            1&~~~0&~~~0&~~~0&~~~0&~~~0\\
                            0&~~~0&~~~0&~~~1&~~~0&~~~0\\
                            0&~~~1&~~~0&~~~0&~~~0&~~~0\\
                            0&~~~0&~~~0&~~~0&~~~1&~~~0\\
                            0&~~~0&~~~1&~~~0&~~~0&~~~0\\
                            0&~~~0&~~~0&~~~0&~~~0&~~~1
                           \end{array}\right]& \!\!\begin{array}{c}
                            (11)\\  (21)\\ (12)\\ (22)\\ (13)\\ (23)
                           \end{array}
\end{array}.
$$
When $m=n$, $W_{[m,n]}$ is denoted by $W_{[m]}$.
\begin{lemma}\cite{cheng2010analysis}
\label{lem2}
Swap matrices have the following properties:
\par
1) $W_{[m,n]}^\mathrm{T}=W_{[m,n]}^{-1}=W_{[n,m]}$;
\par
2) $W_{[pq,\ r]}=(W_{[p,\ r]}\otimes I_q)(I_p\otimes W_{[q,\ r]})$;
\par
3) $W_{[m,\ p]}(A\otimes B)W_{[q,\ n]}=B\otimes A$ for $m\times n$ matrix $A$ and $p\times q$ matrix $B$; \par
4) Let $x$ and $\tilde x$ be the $mn$-dimensional column vectors with each entry $x_{ij}$ arranged by Id($i_1,i_2;m,n$) and Id($i_2,i_1;n,m$), respectively. Then $\tilde x=W_{[m,n]}x$. In particular, for any $m$-dimensional vector $x$ and $n$-dimensional vector $y$, one has $yx=W_{[m,n]}xy$.
\end{lemma}
By Lemma \ref{lem2}, one can easily get
\begin{eqnarray}
\label{eqn4}
~~~~&& (I_{\!k}\otimes W_{\![k]}\!)(W_{\![k]}\otimes I_{\!k})(I_{\!k}\otimes W_{\![k]})\!\nonumber \\ ~~~~&=&(W_{\![k]}\otimes I_{\!k})(I_{\!k}\otimes W_{\![k]})(W_{\![k]}\otimes I_{\!k}).
\end{eqnarray}

\section{Symmetric Games and Testing Conditions}
This section will present a simple matrix approach to test symmetric games.
\begin{definition}\cite{neumann1944theory}
\label{def4}
A game $G\in\mathcal{G}_{[n;k]}$ is called a symmetric game if for any permutation $\sigma\in \mathbf{S}_n$
\begin{equation}
\label{eqn5}
c_i(x_1,\cdots,x_n)=c_{\sigma(i)}(x_{\sigma^{-1}(1)},\cdots,x_{\sigma^{-1}(n)})
\end{equation}
for any $i=1, 2, \cdots, n.$
\end{definition}
By Definition \ref{def4}, to test whether $G\in\mathcal{G}_{[n;k]}$ is symmetric, it seems that (\ref{eqn5}) should be checked for every permutation $\sigma$. But actually it is only needed to check
 (\ref{eqn5}) for the permutations of a generator of $G\in\mathcal{G}_{[n;k]}$. This is due  to the fact that the compound permutation  $\sigma_2\circ\sigma_1$ satisfies (\ref{eqn5}) if both permutations $\sigma_1$ and $\sigma_2$ satisfy (\ref{eqn5}). For the readability, we write the demonstration as follows:
 \par For any given $x_i\in \Delta_k$, $i=1,2,\dots,n$, let $y_i=x_{\sigma_1^{-1}(i)}$. Then
\begin{eqnarray}
\label{eq5}
c_i(x_1,\cdots,x_n)&=&c_{\sigma_1(i)}(x_{\sigma_1^{-1}(1)},\cdots,x_{\sigma_1^{-1}(n)}) \nonumber \\
&=&  c_{\sigma_1(i)}(y_{1},\cdots,y_{n}) \nonumber \\
&=&  c_{\sigma_2\circ\sigma_1(i)}(y_{\sigma_2^{-1}(1)},\cdots,y_{\sigma_2^{-1} (n)}) \nonumber \\
&=&  c_{\sigma_2\circ\sigma_1(i)}(x_{\sigma_1^{-1}(\sigma_2^{-1}(1))},\cdots,x_{\sigma_1^{-1}(\sigma_2^{-1} (n))}),\nonumber
\end{eqnarray}
which implies that $\sigma_2\circ\sigma_1$ satisfies (\ref{eqn5}).

\begin{lemma}
\label{lem2}
\cite{jerrum1985complexity,johnson1963generation}
The set of all the adjacent transpositions $(r,r+1)$, $1\leq r\leq n-1$ is a generator of the symmetric group $\mathbf{S}_n$.
\end{lemma}

In the following, we will use $\mu_r$ to represent the adjacent transposition $(r,r+1)$.
From Definition \ref{def4} and Lemma \ref{lem2}, the following lemma follows:
 \begin{lemma}
\label{lem4}
A game $G\in\mathcal{G}_{[n;k]}$ is a symmetric game if and only if
\begin{equation}
\label{eqn6}
c_i(x_1,\cdots,x_n)=c_{\mu_r(i)}(x_{\mu_r(1)},\cdots,x_{\mu_r(n)})
\end{equation}
for any adjacent transposition $\mu_r$, $1\leq r\leq n-1$, $i=1, 2, \cdots, n.$
\end{lemma}
Now we can use the semi-tensor product method to investigate symmetric games.
\begin{proposition}
\label{prop1}
Consider a finite game $G\in\mathcal{G}_{[n;k]}$. $G$ is a symmetric game if and only if
\begin{equation}
\label{eqn7}
V_i^c=V_{\mu_r(i)}^cT_{\mu_r}, ~\forall i\in N, ~1\leq r\leq n-1,
\end{equation}
where $T_{\mu_r}\!=\!I_{k^{r-1}}\otimes W_{[k]}\otimes I_{k^{n-r-1}}$.
\end{proposition}
\begin{pf}
By (\ref{eqn3}), we rewrite the right hand side of (6) into the matrix form as follows:
\begin{eqnarray}
\label{eqn8}
&&V_{\mu_r(i)}^cx_{\mu_r(1)}x_{\mu_r(2)}\cdots x_{\mu_r(n)}\nonumber \\
&=&V_{\mu_r(i)}^cx_{1}\cdots x_{r-1}(x_{r+1}x_{r})x_{r+2}\cdots x_{n} \nonumber \\
&=& V_{\mu_r(i)}^c T_{\mu_r}x_{1}\cdots x_{n}.
\end{eqnarray}
From (\ref{eqn3}) and (\ref{eqn8}), it follows that (\ref{eqn7}) is equivalent to(\ref{eqn6}). Therefore, by Lemma \ref{lem2}, the proposition is proved. \hfill $\blacksquare$
\end{pf}
Now, we present our main result on testing symmetric games. For the readability of this paper and without loss of generality, we first consider the case of $n=4$.

\begin{theorem}
\label{th1}
Consider $G\in\mathcal{G}_{[4;k]}$. $G$ is a symmetric game if and only if
\begin{equation}
\label{eqn9}
\begin{bmatrix}
I_{{k^4}}          & -T_{\mu_1} & 0 & 0   \\
0 & I_{{k^4}}          &-T_{\mu_2}  & 0   \\
0 & 0 & I_{{k^4}}          &-T_{\mu_{3}}  \\
0 & 0 & 0 & I_{{k^4}}\!-\!T_{\mu_1}\\
0 & 0 & 0  & I_{{k^4}}\!-\!T_{\mu_2}
\end{bmatrix}
(V_G)^\mathrm{T}=0,
\end{equation}
where
\begin{equation}
\label{eqn10}
T_{\mu_1}\!\!=\!W_{[k]}\!\otimes \!I_{k^{2}},\ T_{\mu_2}\!\!=\!I_{k}\!\otimes\! W_{[k]}\!\otimes\! I_{k},\ T_{\mu_3} \!\!=\!I_{k^2}\!\otimes\! W_{[k]},
\end{equation}
and
\begin{equation}
\label{eqn11}
V_G=[V_1^c~V_2^c~V_3^c~V_4^c].
\end{equation}
\end{theorem}
\begin{pf}
First of all, by the properties of Kronecker product and the property (\ref{eqn4}) of swap matrices, we have the properties of $T_{\mu_i}$ as follows:
\begin{equation}
\begin{split}
\label{eqn12}
&T_{\!\mu_i}^2\!\!=\!I_{{k^4}},\ T_{\!\mu_1}T_{\!\mu_3}\!\!=\!T_{\!\mu_3}T_{\!\mu_1},
\\
&T_{\!\mu_1}T_{\!\mu_2}T_{\!\mu_1}\!\!=\!T_{\!\mu_2}T_{\!\mu_1}T_{\!\mu_2},\ T_{\!\mu_2}T_{\!\mu_3}T_{\!\mu_2}\!\!=\!T_{\!\mu_3}T_{\!\mu_2}T_{\!\mu_3}.
\end{split}
\end{equation}
We write (\ref{eqn7}) of Proposition \ref{prop1} for every $r=1,2,3$ as follows:
\begin{eqnarray}
\label{eqn13}
&&V_1^c=V_{2}^cT_{\mu_1},\ V_3^c=V_{3}^cT_{\mu_1}, \ V_4^c=V_{4}^cT_{\mu_1};
 \\
\label{eqn14}
&&V_1^c=V_{1}^cT_{\mu_2},\ V_2^c=V_{3}^cT_{\mu_2}, \ V_4^c=V_{4}^cT_{\mu_2};
\\
\label{eqn15}
&&V_1^c=V_{1}^cT_{\mu_3},\ V_2^c=V_{2}^cT_{\mu_3}, \ V_3^c=V_{4}^cT_{\mu_3}.
\end{eqnarray}
Let
\begin{equation}
\label{eqn16}
A_1\!=\!\begin{bmatrix}
I_{{k^4}}-T_{\mu_2}   \\
I_{{k^4}}-T_{\mu_3}
\end{bmatrix}\!,\
\ A_2=I_{{k^4}}-T_{\mu_3},
\end{equation}
\begin{equation}
\label{eqn17}
A_3=I_{{k^4}}-T_{\mu_1},\
A_4\!=\!\begin{bmatrix}
I_{{k^4}}-T_{\mu_1}   \\
I_{{k^4}}-T_{\mu_2}
\end{bmatrix}.
\end{equation}
Then Eqs. (\ref{eqn13})-(\ref{eqn15}) can be rewritten as
\begin{equation}
\label{eqn18}
\begin{bmatrix}
I_{{k^4}}          & -T_{\mu_1} & 0 & 0   \\
0 & I_{{k^4}}          &-T_{\mu_2}  & 0   \\
0 & 0 & I_{{k^4}}          &-T_{\mu_{3}}  \\
A_1 & 0 & 0 & 0\\
0 & A_2 & 0  & 0\\
0 & 0   & A_3  & 0\\
0 & 0   &  0  & A_4
\end{bmatrix}
(V_G)^\mathrm{T}=0.
\end{equation}
By using an elementary row transformation, one can easily get the equivalent form of (\ref{eqn18}) as follows:
\begin{equation}
\label{eqn19}
\begin{bmatrix}
I_{{k^4}}          & -T_{\mu_1} & 0 & 0   \\
0 & I_{{k^4}}          &-T_{\mu_2}  & 0   \\
0 & 0 & I_{{k^4}}          &-T_{\mu_{3}}  \\
0 & 0 & 0 & \tilde A_3\\
0 & 0 & 0 & A_4
\end{bmatrix}
(V_G)^\mathrm{T}=0,
\end{equation}
where
\begin{eqnarray}
\label{eqn20}
\tilde A_3\!&=&-
\!\begin{bmatrix}
A_1 & 0 & 0  \\
0 & A_2 & 0 \\
0 & 0   & A_3
\end{bmatrix}
\!\!\begin{bmatrix}
I_{{k^4}} & -T_{\mu_1} & 0   \\
0       &  I_{{k^4}}   &-T_{\mu_2}    \\
0       &  0         & I_{{k^4}}
\end{bmatrix}^{\!-1}
\!\!\begin{bmatrix}
 0   \\
 0   \\
-T_{\mu_{3}}
\end{bmatrix} \nonumber \\
&=&
\ \begin{bmatrix}
A_1 & 0 & 0  \\
0 & A_2 & 0 \\
0 & 0   & A_3
\end{bmatrix}
\!\!\begin{bmatrix}
I_{{k^4}} & T_{\mu_1} & \ T_{\mu_1}\!T_{\mu_2}   \\
0       &  I_{{k^4}}   & T_{\mu_2}    \\
0       &  0         & I_{{k^4}}
\end{bmatrix}
\!\!\begin{bmatrix}
 0   \\
 0   \\
T_{\mu_{3}}
\end{bmatrix} \nonumber \\
&=&\
\begin{bmatrix}
A_1T_{\mu_1}\!T_{\mu_2}T_{\mu_{3}}   \\
A_2 T_{\mu_2}T_{\mu_{3}}    \\
A_3T_{\mu_{3}}
\end{bmatrix}.
\end{eqnarray}
Applying a group of elementary row transformations to $\tilde A_3$ yields
\begin{eqnarray}
\label{eqn21}
\hat A_3&=&\begin{bmatrix}
(I_2\otimes T_{\mu_3}\!T_{\mu_2}T_{\mu_{1}})A_1T_{\mu_1}T_{\mu_2}T_{\mu_{3}}   \\
T_{\mu_3}T_{\mu_{2}}A_2 T_{\mu_2}T_{\mu_{3}}    \\
T_{\mu_{3}}A_3T_{\mu_{3}}
\end{bmatrix}\nonumber \\
&=&\begin{bmatrix}
I_{k^4}-T_{\mu_3}\!T_{\mu_2}T_{\mu_{1}}T_{\mu_{2}}T_{\mu_1}T_{\mu_2}T_{\mu_{3}}   \\
I_{{k^4}}-T_{\mu_3}\!T_{\mu_2}T_{\mu_{1}}T_{\mu_{3}}T_{\mu_1}T_{\mu_2}T_{\mu_{3}}   \\
I_{{k^4}}-T_{\mu_3}T_{\mu_{2}}T_{\mu_{3}} T_{\mu_2}T_{\mu_{3}}    \\
I_{{k^4}}-T_{\mu_{3}}T_{\mu_{1}}T_{\mu_{3}}
\end{bmatrix}.
\end{eqnarray}
Using the properties of $T_{\mu_i}$ shown by (\ref{eqn12}), we have
\begin{eqnarray}
\label{eqn22}
&& T_{\mu_{3}}T_{\mu_{1}}T_{\mu_{3}}=T_{\mu_{3}}T_{\mu_{3}}T_{\mu_{1}}=T_{\mu_{1}}; \\
\label{eqn23}
&& T_{\mu_3}(T_{\mu_{2}}T_{\mu_{3}} T_{\mu_2})T_{\mu_{3}}
=  T_{\mu_3}(T_{\mu_{3}}T_{\mu_{2}}T_{\mu_3})T_{\mu_{3}}=T_{\mu_{2}};\\  &&
\label{eqn24}
T_{\mu_3}\!T_{\mu_2}(T_{\mu_{1}}T_{\mu_{3}}T_{\mu_1})T_{\mu_2}T_{\mu_{3}} = T_{\mu_3}\!T_{\mu_2}T_{\mu_{3}}T_{\mu_2}T_{\mu_{3}}
\nonumber  \\
&&= T_{\mu_3}\!(T_{\mu_3}T_{\mu_{2}}T_{\mu_3})T_{\mu_{3}}=T_{\mu_{2}};\\ && T_{\mu_3}\!T_{\mu_2}(T_{\mu_{1}}T_{\mu_{2}}T_{\mu_1})T_{\mu_2}T_{\mu_{3}} =T_{\mu_3}\!T_{\mu_2}(T_{\mu_{2}}T_{\mu_{1}}T_{\mu_2})T_{\mu_2}T_{\mu_{3}}\nonumber \\
\label{eqn25} &&= T_{\mu_3}T_{\mu_{1}}T_{\mu_{3}}=T_{\mu_{1}}.
\end{eqnarray}
So, (\ref{eqn19}) is equivalent to (\ref{eqn9}). The proof is complete. \hfill $\blacksquare$
\end{pf}

By Theorem 1, symmetric games have been clearly described by a simple form. Then the problem is how to simply determine the dimension and bases of the symmetric game subspace. From (\ref{eqn9}), we see that the key to solve the problem is the following linear equations:
\begin{equation}
\label{eqn26}
\left[\begin{array}{c}
I_{k^3}-W_{[k]}\!\otimes \!I_{k}\\
I_{k^3}-I_{k}\!\otimes\! W_{[k]}
\end{array}\right]x=0,
\end{equation}
where $x$ is the $k^3$-dimensional unknown vector. For the linear equation (\ref{eqn26}), we give a lemma as follows:
\begin{lemma}
\label{lem5}
The dimension of the solution space of linear equations (\ref{eqn26}) is $C_{k+2}^3=\frac{1}{6}k(k+1)(k+2)$.
\end{lemma}
\begin{pf}
Let $x=(x_{pqr})$ be a $k^3$-dimensional column vector arranged
by the ordered multi-index Id$(i_1,i_2,i_3;k,k,k)$, that is,
\begin{equation}
\label{eqn27}
x\!=\!(x_{111},\ldots, x_{11k},x_{121},\ldots, x_{12k},\ldots, x_{kk1},\ldots, x_{kkk})^\mathrm{T}.
\end{equation}
Then, by the property of $W_{[k]}$, $x$ is a solution of (\ref{eqn26}) if and only if
\begin{equation}
\label{eqn28}
x_{pqr}=x_{qpr}, \ \ x_{pqr}=x_{prq}, \ \ \forall \ 1\leq p,q,r\leq k,
\end{equation}
which is equivalent to
\begin{equation}
\label{eqn29}
x_{pqr}=x_{\tau (pqr)},\  \forall \tau\in \ \mathbf{S}_3.
\end{equation}
So all the free variables of the linear equations are
\begin{equation}
\label{eqn30}
x_{pqr},\  \forall \ 1\leq p\leq q \leq r\leq k,
\end{equation}
whose number is $C_{k+2}^3$ by Theorem 2.5.1 of \cite{brualdi1977introductory} on counting number of combinations of a multiset. \hfill $\Box$
\end{pf}

The proof of Lemma \ref{lem5} actually provides an effective approach to construct a group of bases of the solution space $\mathcal{X}_3$ of (\ref{eqn26}). Based on the bases, we also can easily get the minimum number of equivalent equations to (\ref{eqn26}). See the details as follows:
\par
For every given repeatable combination $pqr$ $(1\leq p\leq q\leq r\leq k)$, denote by $P_{pqr}$ the set composed of all the repeatable permutation of $pqr$. For example, $P_{111}=\{111\}$, $P_{112}=\{112, 121, 211\}$, $P_{123}=\{123, 231, 312, 132, 321,213\}$.
\begin{lemma}
\label{lem6}
For every combination $pqr$ $(1\leq p\leq q\leq r\leq k)$, define a vector $\eta^{pqr}=x$ with the form (\ref{eqn27}) by
$$
x_{abc}=\left\{\begin{array}{l}
             1, \ \  abc\in P_{pqr},\\
             0, \ \ \mbox{otherwise}.
             \end{array}\right.
$$
Then the set of $\eta^{pqr}$ $(1\leq p\leq q\leq r\leq k)$ is a group of bases of the solution space $\mathcal{X}_3$ of (\ref{eqn26}).
For every ${pqr}$ $(1\leq p\leq q\leq r\leq k)$ that does not satisfy $p=q=r$, we define $|P_{pqr}|-1$ number of vectors
$\zeta_{pqr}^{uvw}=x$ $(uvw\in P_{pqr}, uvw\ne {pqr})$ by
$$
x_{abc}=\left\{\begin{array}{l}
             1, \ \  abc=pqr,\\
             -1,\ \ abc=uvw,\\
             0, \ \ \mbox{otherwise}.
             \end{array}\right.
$$
Then the set of $\zeta_{pqr}^{uvw}$ $(uvw\in P_{pqr}, uvw\ne {pqr})$ is a group of bases of the orthogonal complementary space $\mathcal{X}_3^{\perp}$.
Denote by $M_\mathcal{W}$ the matrix composed by a group of bases of the subspace  $\mathcal{W}$. Then the linear system  (\ref{eqn26}) is equivalent to $M_{\mathcal{X}_3^\perp}^\mathrm{T}x=0$.
\end{lemma}
\begin{pf}
From the equivalent equations (\ref{eqn29}) and the free variables shown by (\ref{eqn30}), it follows that the set of $\eta^{pqr}$ $(1\leq p\leq q\leq r\leq k)$ is a group of bases of the solution space $\mathcal{X}_3$. By the construction of $\zeta_{pqr}^{uvw}$, it is straightforward to see that each $\zeta_{pqr}^{uvw}$ is orthogonal to $\mathcal{X}_3$. Since the total number of
$\zeta_{pqr}^{uvw}$ is $\sum_{1\leq p\leq q\leq r\leq k}(|P_{pqr}|-1)
=\sum_{1\leq p\leq q\leq r\leq k}|P_{pqr}|-C_{k+2}^3
=k^3-C_{k+2}^3=k^3-\mbox{dim}(\mathcal{X}_3)$,
we conclude that the set of $\zeta_{pqr}^{uvw}$ $(uvw\in P_{pqr}, uvw\ne {pqr})$ is a group of bases of $\mathcal{X}_3^{\perp}$.
\end{pf}
For example, when $k=2$, we have
$$
M_{\mathcal{X}_3}\!\!=\!\!
\left[\begin{array}{cccc}
                            1&~0&~0&~0\\ 
                            0&~1&~0&~0\\ 
                            0&~1&~0&~0\\ 
                            0&~0&~1&~0\\ 
                            0&~1&~0&~0\\ 
                            0&~0&~1&~0 \\
                            0&~0&~1&~0 \\
                            0&~0&~0&~1 \\
                            \end{array}\right]\!\!,\
M_{\mathcal{X}_3^{\perp}}\!\!=
\!\!\left[\begin{array}{cccc}
                            0& 0& 0& 0\\ 
                            1& 1& 0& 0\\ 
                           -1& 0& 0& 0\\ 
                            0& 0& 1& 1\\ 
                            0&-1& 0& 0\\ 
                            0& 0&-1& 0 \\
                            0& 0& 0&-1 \\
                            0& 0& 0& 0 \\
                            \end{array}\right]\ \ \begin{array}{c}
                            (111)\\  (112)\\ (121)\\ (122)\\ (211)\\  (212)\\ (221)\\ (222)
                           \end{array}
$$
Based on Theorem \ref{th1}, Lemma \ref{lem5} and Lemma {lem6}, one can investigate the dimension and bases of the  symmetric game subspace.
\begin{theorem}
\label{thm2}
The dimension of the symmetric game subspace $\mathcal{S}_{[4;k]}$ is $kC_{k+2}^3$. A group of bases of $\mathcal{S}_{[4;k]}$ is composed of the columns of matrix
\begin{equation}
\label{eqn31}
\begin{bmatrix}
W_{[k^3,k]}\\
I_{k}\otimes W_{[k^2,k]}\\
I_{k^2}\otimes W_{[k]} \\
I_{k^4}
\end{bmatrix}
(M_{\mathcal{X}_3}\otimes I_k),
\end{equation}
where $M_{\mathcal{X}_3}$ is composed of a bases of linear equations (\ref{eqn26}). Moreover, the linear equations with the minimum numbers to test symmetric games in $\mathcal{G}_{[4;k]}$ are
\begin{equation}
\label{eqn32}
\begin{bmatrix}
I_{{k^4}}          & -T_{\mu_1} & 0 & 0   \\
0 & I_{{k^4}}          &-T_{\mu_2}  & 0   \\
0 & 0 & I_{{k^4}}          &-T_{\mu_{3}}  \\
0 & 0 & 0 & M_{\mathcal{X}_3^\bot}^\mathrm{T}\otimes I_k
\end{bmatrix}
(V_G)^\mathrm{T}=0.
\end{equation}
\end{theorem}
\begin{pf}
%
It is easy to see that
\begin{equation}
\label{eqn33}
\begin{bmatrix}
 I_{{k^4}}\!-\!T_{\mu_1}\\
 I_{{k^4}}\!-\!T_{\mu_2}
\end{bmatrix}
=\begin{bmatrix}
I_{k^3}-W_{[k]}\!\otimes \!I_{k}\\
I_{k^3}-I_{k}\!\otimes\! W_{[k]}
\end{bmatrix}\otimes I_k.
\end{equation}
By Theorem \ref{th1} and Lemma \ref{lem5}, we can easily get the dimension of the symmetric game subspace $\mathcal{S}_{[4;k]}$ is $kC_{k+2}^3$.
Using $M_{\mathcal{X}_3}$ composed of the bases of (\ref{eqn26}), we get a group of bases of the solution space of (\ref{eqn9}):
\begin{equation}
\label{eqn34}
\begin{bmatrix}
T_{\mu_1}T_{\mu_2}T_{\mu_3}(M_{\mathcal{X}_3}\otimes I_k)\\
T_{\mu_2}T_{\mu_3}(M_{\mathcal{X}_3}\otimes I_k)\\
T_{\mu_3}(M_{\mathcal{X}_3}\otimes I_k) \\
M_{\mathcal{X}_3}\otimes I_k
\end{bmatrix}.
\end{equation}
By the properties of swap matrices, we have that
$$
T_{\mu_1}T_{\mu_2}T_{\mu_3}=W_{[k^3,k]},\ \ T_{\mu_2}T_{\mu_3}=I_k\otimes W_{[k^2,k]}.
$$
So, (\ref{eqn34}) is just (\ref{eqn31}).
Moreover, from (\ref{eqn33}) and Lemma \ref{lem6}, it follows that (\ref{eqn9}) is equivalent to (\ref{eqn32}).
Since the coefficient matrix of (\ref{eqn32}) has a full row rank, the testing equations have the minimum number.
\hfill $\blacksquare$
\end{pf}

 In the rest of this paper, we consider the general finite game
$G\in\mathcal{G}_{[n;k]}$.
\begin{theorem}
\label{th3}
Consider $G\in\mathcal{G}_{[n;k]}$. $G$ is a symmetric game if and only if
\begin{equation}
\label{eqn35}
\begin{bmatrix}
I_{k^n} & -T_{\mu_1} & &  &&\\
& I_{k^n} &-T_{\mu_2}&  &&\\
 &  & \ddots & \ddots & &  \\
& &&   I_{k^n} & -T_{\mu_{n-1}} \\
& & &   & I_{k^n}\!-\!T_{\mu_1}\\
& & &   & I_{k^n}\!-\!T_{\mu_2}\\
 & & &  & \vdots  \\
& & & & I_{k^n}\!-\!T_{\mu_{n-2}}
\end{bmatrix}
V_G^{\rm T}=0,
\end{equation}
where
\begin{equation}
\label{eqn36}
T_{\mu_r}=I_{k^{r-1}}\otimes W_{[k]}\otimes I_{k^{n-r-1}}, \ 1\leq r\leq n-1\\
\end{equation}
and
\begin{equation}
V_G=[V_1^c~V_2^c~\cdots~V_n^c].
\end{equation}
\end{theorem}
\begin{pf}
According to Proposition $\ref{prop1}$, $G\in\mathcal{G}_{[n;k]}$ is a symmetric game if and only if
\begin{equation}
\label{eqn37}
\begin{bmatrix}
I_{k^n} & -T_{\mu_1} & &  &&\\
& I_{k^n} &-T_{\mu_2}&  &&\\
 &  & \ddots & \ddots & &  \\
& &&   I_{k^n} & -T_{\mu_{n-1}} \\
A_1& &  & & \\
& A_2  & & \\
 &  &\ddots & &   \\
 &  & & A_{n-1}& \\
  & & & &A_{n}
\end{bmatrix}
V_G^{\rm T}=0,
\end{equation}
where
\begin{equation}
\label{eqn38}
A_1=\begin{bmatrix}
        I_{k^n}-T_{\mu_2} \\
        I_{k^n}-T_{\mu_3} \\
        \vdots\\
        I_{k^n}-T_{\mu_{n-1}}
    \end{bmatrix},\ A_2=\begin{bmatrix}
        I_{k^n}-T_{\mu_3} \\
        I_{k^n}-T_{\mu_4} \\
        \vdots\\
        I_{k^n}-T_{\mu_{n-1}}
    \end{bmatrix},
\end{equation}
\begin{equation}
\label{eqn39}
A_r\!=\!\!\begin{bmatrix}
        I_{k^n}\!-\!T_{\mu_1} \\
        \vdots\\
        I_{k^n}\!-\!T_{\mu_{r-2}} \\
        I_{k^n}\!-\!T_{\mu_{r+1}} \\
        \vdots\\
        I_{k^n}\!-\!T_{\mu_{n-1}}
    \end{bmatrix}\!(\ 3\!\leq\! r\!\leq \!n-2), \
\end{equation}
\begin{equation}
\label{eqn40}
A_{n-1}\!=\!\!\begin{bmatrix}
        I_{k^n}\!-\!T_{\mu_1} \\
        I_{k^n}\!-\!T_{\mu_2} \\
        \vdots\\
        I_{k^n}\!-\!T_{\mu_{n-3}} \\
    \end{bmatrix},
A_n\!=\!\!\begin{bmatrix}
        I_{k^n}\!-\!T_{\mu_1} \\
        I_{k^n}\!-\!T_{\mu_2} \\
        \vdots\\
        I_{k^n}\!-\!T_{\mu_{n-2}}
    \end{bmatrix}.
\end{equation}

Let $F_1=I_{n-2}\otimes(T_{\mu_{n-1}}T_{\mu_{n-2}}\cdots T_{\mu_1})$
and $F_r=I_{n-3}\otimes(T_{\mu_{n-1}}T_{\mu_{n-2}}\cdots T_{\mu_r})$ for every $2\leq r\leq n-1$. Then a series of elementary row transformations to $(\ref{eqn37})$ results in
\begin{equation}
\label{eqn41}
\begin{bmatrix}
I_{k^n} & -T_{\mu_1} & &  &&\\
& I_{k^n} &-T_{\mu_2}&  &&\\
 &  & \ddots & \ddots & &  \\
& &&   I_{k^n} & -T_{\mu_{n-1}} \\
& &  & & \!\!F_1A_1T_{\mu_1}T_{\mu_2}\!\!\cdots \!\!T_{\mu_{n-1}}\\
& &  & & \!\!F_2A_2T_{\mu_2}T_{\mu_3}\!\!\cdots \!\!T_{\mu_{n-1}}\\
 &  & & &  \vdots \\
 & &  & & \!\!F_{n-1}A_{n-1}T_{\mu_{n-1}}& \\
  & & & &A_{n}\!\!
\end{bmatrix}
\!\!V_G^{\rm T}\!=\!0.
\end{equation}
From the properties of swap matrix $W_{[k]}$, it follows that
\begin{equation}
\label{eqn42}
T_{\mu_{n-1}}\cdots T_{\mu_r}T_{\mu_i}T_{\mu_r}\!\cdots \!T_{\mu_{n-1}}\!=\!
\left\{\!\!\begin{array}{l}
    T_{\mu_i},\! ~~~~~\forall \ 1\!\leq \!i\!\leq\! r\!-\!2,\\
    T_{\mu_{i-1}}, ~~\forall\ r\!+\!1\!\leq\! i\!<\!n.
\end{array}
\right.
\end{equation}
Substituting (\ref{eqn42}) into (\ref{eqn41}) yields the equivalent linear equations (\ref{eqn35}). Thus the proof is complete. \hfill$\blacksquare$
\end{pf}
To get the dimension and bases of a general symmetric game subspace, we consider the following linear equations:
\begin{equation}
\label{eqn43}
\begin{bmatrix}
        I_{k^{n-1}}-W_{[k]}\otimes I_{k^{n-3}} \\
        I_{k^{n-1}}-I_k\otimes W_{[k]}\otimes I_{k^{n-4}} \\
        \vdots\\
       I_{k^{n-1}}-  I_{k^{n-3}}\otimes W_{[k]}
    \end{bmatrix}x=0.
\end{equation}
Let $x=(x_{p_1p_2\cdots p_{n-1}})$ be a $k^{n-1}$-dimensional column vector arranged
by the ordered multi-index Id$(i_1,i_2,\ldots,i_{n-1};k,k,\ldots,k)$.
Then, by the property of $W_{[k]}$, $x$ is a solution of (\ref{eqn43}) if and only if
\begin{equation}
\label{eqn44}
x_{p_1p_2\cdots p_{n-1}}=x_{\mu_r(p_1p_2\cdots p_{n-1})}
\end{equation}
for any adjacent transposition $\mu_r$, $1\leq r\leq n-2$ and  $1\leq p_1,p_2,\ldots,p_{n-1}\leq k$,
which is equivalent to
\begin{equation}
\label{eqn45}
x_{p_1p_2\cdots p_{n-1}}=x_{\sigma(p_1p_2\cdots p_{n-1})} \end{equation}
for any $\sigma\in \mathbf{S}_{n-1}$ by Lemma \ref{lem2}. Therefore, the free variables of the linear equations can be chosen as
\begin{equation}
\label{eqn46}
x_{p_1p_2\cdots p_{n-1}}\ \ (1\!\leq \!p_1 \!\leq \!p_2 \!\leq \!\cdots \!\leq \!p_{n-1}\!\leq k).
\end{equation}

\begin{lemma}
\label{lem7}
The dimension of the solution space of linear equations (\ref{eqn43}) is $C_{k+n-2}^{n-1}$.
For every repeatable combination $p_1p_2\cdots p_{n-1}$ $(1\!\leq \!p_1 \!\leq \!p_2 \!\leq \!\cdots \!\leq \!p_{n-1}\!\leq k)$,
let $P_{p_1p_2\cdots p_{n-1}}$ be the set composed of all the repeatable permutations of $p_1p_2\cdots p_{n-1}$.
Define a vector $\eta^{p_1p_2\cdots p_{n-1}}=x$ by
$$
x_{a_1a_2\cdots a_{n-1}}=\left\{\begin{array}{l}
             1, \ \  a_1a_2\cdots a_{n-1}\in P_{p_1p_2\cdots p_{n-1}},\\
             0, \ \ \mbox{otherwise}.
             \end{array}\right.
$$
Then the set of $\eta^{p_1p_2\cdots p_{n-1}}=x$ $(1\!\leq \!p_1 \!\leq \!p_2 \!\leq \!\cdots \!\leq \!p_{n-1}\!\leq k)$ is a group of bases of the solution space $\mathcal{X}_{n-1}$ of (\ref{eqn43}).
Define
$\zeta_{p_1p_2\cdots p_{n-1}}^{q_1q_2\cdots q_{n-1}}=x$ $(q_1q_2\cdots q_{n-1}\in P_{p_1p_2\cdots p_{n-1}}, q_1q_2\cdots q_{n-1}\ne p_1p_2\cdots p_{n-1})$ by
$$
x_{a_1a_2\cdots a_n}=\left\{\begin{array}{l}
             1, \ \  a_1a_2\cdots a_n=p_1p_2\cdots p_{n-1},\\
             -1,\ \ a_1a_2\cdots a_n=q_1q_2\cdots q_{n-1},\\
             0, \ \ \mbox{otherwise}.
             \end{array}\right.
$$
Then $M_{\mathcal{X}_{n-1}}$ is composed of $\eta^{p_1p_2\cdots p_{n-1}}$ $(1\!\leq \!p_1 \!\leq \!p_2 \!\leq \!\cdots \!\leq \!p_{n-1}\!\leq k)$, and $M_{\mathcal{X}_{n-1}^\perp}$ is composed of $\zeta_{p_1p_2\cdots p_{n-1}}^{q_1q_2\cdots q_{n-1}}$ $(q_1q_2\cdots q_{n-1}\in P_{p_1p_2\cdots p_{n-1}}, q_1q_2\cdots q_{n-1}\ne p_1p_2\cdots p_{n-1})$.
\end{lemma}

\begin{theorem}
\label{thm4}
The dimension of the symmetric game subspace $\mathcal{S}_{[n;k]}$ is $kC_{k+n-2}^{n-1}$. A group of bases of $\mathcal{S}_{[n;k]}$ is composed of the columns of matrix
\begin{equation}
\label{eqn47}
\begin{bmatrix}
W_{[k^{n-1},k]}\\
I_{k}\otimes W_{[k^{n-2},k]}\\
I_{k^2}\otimes W_{[k^{n-3},k]}\\
 \vdots\\
I_{k^{n-2}}\otimes W_{[k]} \\
I_{k^n}
\end{bmatrix}
(M_{\mathcal{X}_{n-1}}\otimes I_k),
\end{equation}
where $M_{\mathcal{X}_{n-1}}$ is composed of a bases of linear equations (\ref{eqn43}). Moreover, the linear equations with the minimum numbers to test symmetric games in $\mathcal{G}_{[n;k]}$  are
\begin{equation}
\label{eqn48}
\begin{bmatrix}
I_{k^n} & -T_{\mu_1} & &  &&\\
& I_{k^n} &-T_{\mu_2}&  &&\\
 &  & \ddots & \ddots & &  \\
& &&   I_{k^n} & -T_{\mu_{n-1}} \\
& & &   & M_{\mathcal{X}_{n-1}^\perp}^\mathrm{T}\!\!\!\!\!\otimes I_k
\end{bmatrix}
V_G^{\rm T}=0.
\end{equation}
\end{theorem}
\begin{pf}
%
It is easy to see that
\begin{equation}
\label{eqn49}
\begin{bmatrix}
 I_{k^n}\!-\!T_{\mu_1}\\
 I_{k^n}\!-\!T_{\mu_2}\\
 \vdots  \\
I_{k^n}\!-\!T_{\mu_{n-2}}
\end{bmatrix}
=\begin{bmatrix}
        I_{k^{n-1}}-W_{[k]}\otimes I_{k^{n-3}} \\
        I_{k^{n-1}}-I_k\otimes W_{[k]}\otimes I_{k^{n-4}} \\
        \vdots\\
       I_{k^{n-1}}-  I_{k^{n-3}}\otimes W_{[k]}
    \end{bmatrix}\otimes I_k.
\end{equation}
By Theorem \ref{th3} and Lemma \ref{lem7}, we can easily get the dimension of the symmetric game subspace $\mathcal{S}_{[n;k]}$ is $kC_{k+n-2}^{n-1}$.
Using $M_{\mathcal{X}_{n-1}}$ composed of the bases of (\ref{eqn43}), we get a group of bases of the solution space of (\ref{eqn35}):
\begin{equation}
\label{eqn50}
\begin{bmatrix}
T_{\mu_1}T_{\mu_2}\cdots T_{\mu_{n-1}}(M_{\mathcal{X}_{n-1}}\otimes I_k)\\
T_{\mu_2}T_{\mu_{n-1}}(M_{\mathcal{X}_{n-1}}\otimes I_k)\\
\vdots\\
T_{\mu_{n-1}}(M_{\mathcal{X}_{n-1}}\otimes I_k) \\
M_{\mathcal{X}_{n-1}}\otimes I_k
\end{bmatrix}.
\end{equation}
By the properties of swap matrices, we have that
$$
T_{\mu_s}T_{\mu_{s+1}}\cdots T_{\mu_{n-1}}=I_k^{s-1}\otimes W_{[k^{n-s},k]}
$$
for each $1\leq s\leq n-1$. So, (\ref{eqn50}) is just (\ref{eqn47}).
Moreover, from (\ref{eqn49}), it follows that (\ref{eqn35}) is equivalent to (\ref{eqn48}).
Since the coefficient matrix of (\ref{eqn48}) has a full row rank, the testing equations have the minimum number.
\hfill $\blacksquare$
\end{pf}

\begin{remark}
In Theorem 19.2.1 of \cite{cheng2016mapping}, the dimension of the symmetric game subspace is obtained by using the concept of strategy multiplicity vector of \cite{cao2018symmetric}. However, Theorem \ref{thm4} gives the dimension of a symmetric game subspace via a purely  algebraic approach without using the strategy multiplicity vector, which makes the proof greatly simplified.
\end{remark}

\begin{remark}
In \cite{hao2018on} and \cite{li2019symmetry}, construction algorithms of a group of bases for the symmetric game subspace are given by using the classification of strategy profiles. Different from \cite{hao2018on} and \cite{li2019symmetry}, Theorem \ref{thm4} directly constructs a group of bases of a symmetric game subspace from a group of bases of the solution space of linear equations (\ref{eqn43}). In addition, the testing equations for symmetric games with the minimum number is easily obtained.
\end{remark}

\begin{example}
Consider $G\in\mathcal{G}_{[3;2]}$.
$$
M_{\mathcal{X}_2}\!\!=\!\!
\left[\begin{array}{cccc}
                            1&~0&~0\\ 
                            0&~1&~0\\ 
                            0&~1&~0\\ 
                            0&~0&~1\\ 
                            \end{array}\right]\!\!,\
M_{\mathcal{X}_2^{\perp}}\!\!=
\!\!\left[\begin{array}{cccc}
                            0\\ 
                            1\\ 
                           -1\\ 
                            0\\ 
                            \end{array}\right]\ \ \begin{array}{c}
                            (11)\\  (12)\\ (21)\\ (22)
                           \end{array}.
$$
$$T_{\mu_1}=W_{[2]}\otimes I_2=\delta_8[1~2~5~6~3~4~7~8];$$
$$T_{\mu_2}=I_2\otimes W_{[2]} =\delta_8[1~3~2~4~5~7~6~8].$$
$G$ is a symmetric game if and only if
\begin{equation}
\begin{bmatrix}
I_{8} & -T_{\mu_1} \\
& I_{8} &-T_{\mu_2}\\
& &   M_{\mathcal{X}_{2}^\perp}^\mathrm{T}\otimes I_2
\end{bmatrix}
V_G^{\rm T}=0.
\end{equation}
Let
$$V_{3}^c=[x_1~x_2~x_3~x_4~x_5~x_6~x_7~x_8].$$
According to $(M_{\mathcal{X}_{2}^\perp}^\mathrm{T}\otimes I_2)(V_3^c)^\mathrm{T}=0$, we have
$$x_3=x_5,~ x_4=x_6.$$
Therefore,
$$V_3^c=[a_1~a_2~a_3~a_4~a_3~a_4~a_5~a_6]$$
where $a_1, a_2, a_3, a_4, a_5, a_6\in\mathbb{R}$.
According to $(V_2^c)^\mathrm{T}=T_{\mu_2}(V_{3}^c)^\mathrm{T}$, $(V_1^c)^\mathrm{T}=T_{\mu_1}(V_{2}^c)^\mathrm{T}$,
$$V_2^c=[a_1~a_3~a_2~a_4~a_3~a_5~a_4~a_6],$$
$$V_1^c=[a_1~a_3~a_3~a_5~a_2~a_4~a_4~a_6].$$
Therefore, the game is a symmetric game if and only if its payoff matrix is in the form
\begin{gather*}
\begin{bmatrix}
a_1 & a_3 & a_3 & a_5 & a_2 & a_4 & a_4 & a_6\\
a_1 & a_3 & a_2 & a_4 & a_3 & a_5 & a_4 & a_6\\
a_1 & a_2 & a_3 & a_4 & a_3 & a_4 & a_5 & a_6\\
\end{bmatrix}.
\end{gather*}
\end{example}

\begin{example}
For any $G\in\mathcal{G}_{[4;2]}$, we let
\begin{gather*}
\begin{split}
T_{\mu_1}=&W_{[2]}\otimes I_4\\
=&\delta_{16}[1~2~3~4~9~10~11~12~5~6~7~8~13~14~15~16],
\end{split}
\end{gather*}
\begin{gather*}
\begin{split}
T_{\mu_2}=&I_2\otimes W_{[2]}\otimes I_2 \\
=&\delta_{16}[1~2~5~6~3~4~7~8~9~10~13~14~11~12~15~16],
\end{split}
\end{gather*}
\begin{gather*}
\begin{split}
T_{\mu_3}=&I_4\otimes W_{[2]}\\
=&\delta_{16}[1~3~2~4~5~7~6~8~9~11~10~12~13~15~14~16].
\end{split}
\end{gather*}
By (\ref{eqn32}), we have
$$(M_{\mathcal{X}_3^\bot}^\mathrm{T}\otimes I_2)(V_4^c)^\mathrm{T}=0,$$
$$(V_3^c)^\mathrm{T}=T_{\mu_3}(V_{4}^c)^\mathrm{T},$$
$$(V_2^c)^\mathrm{T}=T_{\mu_2}(V_{3}^c)^\mathrm{T},$$
$$(V_1^c)^\mathrm{T}=T_{\mu_1}(V_{2}^c)^\mathrm{T}.$$

 Let $V_4^c=(x_1, x_2,\cdots, x_{16})$. According to $(M_{\mathcal{X}_3^\bot}^\mathrm{T}\otimes I_2)(V_4^c)^\mathrm{T}=0$,
we have
\begin{align*}
x_3&=x_5=x_9,\\
x_4&=x_6=x_{10}, \\
x_7&=x_{11}=x_{13},\\
x_8&=x_{12}=x_{14}.
\end{align*}
\end{example}
Then,
\begin{align*}
V_4^c=[b_1~b_2~b_3~b_4~b_3~b_4~b_5~b_6~b_3~b_4~b_5~b_6~b_5~b_6~b_7~b_8],
\end{align*}
where $b_1, b_2, \cdots, b_8\in\mathbb{R}$.
\begin{align*}
V_3^c&=V_4^c T_{\mu_3}\\
&=[b_1~b_3~b_2~b_4~b_3~b_5~b_4~b_6~b_3~b_5~b_4~b_6~b_5~b_7~b_6~b_8];
\end{align*}
\begin{align*}
V_2^c&=V_3^cT_{\mu_2}\\
&=[b_1~b_3~b_3~b_5~b_2~b_4~b_4~b_6~b_3~b_5~b_5~b_7~b_4~b_6~b_6~b_8];
\end{align*}
\begin{align*}
V_1^c&=V_2^cT_{\mu_1}\\
&=[b_1~b_3~b_3~b_5~b_3~b_5~b_5~b_7~b_2~b_4~b_4~b_6~b_4~b_6~b_6~b_8].
\end{align*}
Therefore, the game is a symmetric game if and only if its payoff matrix is in the form
\begin{gather*}
\begin{bmatrix}
b_1~b_3~b_3~b_5~b_3~b_5~b_5~b_7~b_2~b_4~b_4~b_6~b_4~b_6~b_6~b_8\\
b_1~b_3~b_3~b_5~b_2~b_4~b_4~b_6~b_3~b_5~b_5~b_7~b_4~b_6~b_6~b_8\\
b_1~b_3~b_2~b_4~b_3~b_5~b_4~b_6~b_3~b_5~b_4~b_6~b_5~b_7~b_6~b_8\\
b_1~b_2~b_3~b_4~b_3~b_4~b_5~b_6~b_3~b_4~b_5~b_6~b_5~b_6~b_7~b_8\\
\end{bmatrix}.
\end{gather*}

\section{Conclusion}
In this paper, new equivalent conditions for testing symmetric games are obtained via a matrix semi-tensor product method with adjacent transpositions. Compared with the existing literature, this paper provides simple method with purely algebraic operations. Based  on the new equivalent conditions for symmetric games, the dimension of a symmetric game subspace has been directly calculated, and a group of bases of the symmetric game subspace has been easily constructed. In addition, the testing equations with the minimum number have been derived.
\begin{ack}                               
The research work of X. Liu is supprted in part by Natural Science Foundation of Shandong Province of China under Grant ZR2020QA028, and in part by Doctoral Research
Foundation of Weifang University under Grant 2019BS03.
The research work of T. Li and J. Zhu is supported in part by National Natural Science Foundation
(NNSF) of China under Grants 62103194 and 61673012, respectively.  
\end{ack}

\nocite{candogan2011flows}
\nocite{li2019symmetry}
\nocite{hao2018on}
\nocite{Cheng2016}
\nocite{ZHANG2018862}
\nocite{plan2017symmetric}
\nocite{cheng2012introduction}
\nocite{nash1951non}
\nocite{cheng2015modeling}
\nocite{monderer1996potential}
\nocite{hefti2017equilibria}
\nocite{gibbons1992primer}

\bibliographystyle{plain}        
\bibliography{REF}           





\end{document}